# PicoDomain: A Compact High-Fidelity Cybersecurity Dataset

Craig Laprade   Benjamin Bowman   H. Howie Huang
George Washington University

*Abstract -* Analysis of cyber relevant data has become an area of increasing focus. As larger percentages of businesses and governments begin to understand the implications of cyberattacks, the impetus for better cybersecurity solutions has increased. Unfortunately, current cybersecurity datasets either offer no ground truth or do so with anonymized data. The former leads to a quandary when verifying results and the latter can remove valuable information. Additionally, most existing datasets are large enough to make them unwieldy during prototype development. In this paper we have developed the PicoDomain dataset, a compact high-fidelity collection of Zeek logs from a realistic intrusion using relevant Tools, Techniques, and Procedures. While simulated on a small-scale network, this dataset consists of traffic typical of an enterprise network, which can be utilized for rapid validation and iterative development of analytics platforms. We have validated this dataset using traditional statistical analysis and off-the-shelf Machine Learning techniques.

## I. Introduction

Cybersecurity is a burgeoning field of research. Coupled with recent advances in Machine Learning (ML), a scenario where cyber defenses are bolstered by semi-autonomous systems [1] is beginning to become a reality. In this scenario, platforms trained on the Tactics, Techniques, and Procedures (TTPs), and goals of adversaries would be able to detect the abstracted behaviors of an intruder and react, even when the intruder was using legitimate access methods and masquerading near perfectly as a normal user. This future will require training these platforms on large volumes of cyber relevant data with known ground truth. As a result, this presents a challenge with currently available datasets. Large datasets are often required to train effective ML based detectors, but the amount of time and resources it would take to rapidly iterate an analytic over these large datasets is prohibitive. Additionally, currently available large datasets either offer no ground truth (i.e. ISP level Netflow) or offer ground truth but are heavily anonymized (i.e. LANL 2015) [2]. The former leads to a quandary when verifying results and the latter can remove valuable information from the raw data.

In this work, we believe there is a critical need for a tailored dataset that contains representative traffic from the modern enterprise environment, documented ground truth, and is small enough to use in the iterative development of analytics. To this end, we have developed the **PicoDomain** dataset, which is comprised of Zeek [3] and adversary activity logs over a three day campaign utilizing relevant TTPs that a modern attacker would employ against a target. The campaign follows the Mandiant Attack Lifecycle (MAL) [4] from an initial compromise through the execution of mission goals. To demonstrate the benefit of this dataset, we evaluated both statistical analysis and mature ML techniques for the identification of the adversary activity and showed that it is well suited for a variety of cybersecurity analysis applications. The dataset is hosted at https://github.com/iHeartGraph/PicoDomain

This rest of the paper is organized as follows: It starts with the background in Section II and motivation behind the construction of the dataset in Section III, moves into the environment in Section IV and the adversary actions in Section V, discusses the data collection techniques in section VI, and then concludes with empirical observations in Section VII and proposed future work in Section VIII.

## II. Background

Several outstanding projects already exist to provide cybersecurity data to researchers; however, when assessing them to evaluate ongoing ML research they all had a drawback that limited their usefulness for unsupervised ML approaches. In the following, we will discuss several representative projects. Note that we do not intend to denigrate these resources, most were created without this use in mind and perform exceptionally well for their intended purposes which is often to either train a human analysis or to train a signature, heuristic, or ML based detector that detects a singular TTP.

MITRE Caldera [5] is an automated adversary emulation system, built on the MITRE ATT&CK framework. This platform produces highly detailed artifacts of adversary behavior on systems. It does this using a pre-installed agent that communicates to a Command & Control (C2) server. Although the actual adversarial artifacts are high fidelity, the adjacency to the agent in an execution timeline and the agent C2 traffic degrade the usefulness of produced logs for generalized ML training. On the other hand, Atomic Red Team [6] allows every security team to test their controls by executing simple "atomic tests" that exercise the same techniques used by adversaries (all mapped to MITRE's ATT&CK framework). Atomic Red Team predates Caldera and is one of the principle frameworks referenced in its creation. Atomic Red Team works for small "atomic tests" but is ill suited for testing a platform to sniff out all phases of the kill chain in a dynamic environment.

The LANL Comprehensive Multi-Source Cyber-Security Events Dataset [2] is a comprehensive and diverse dataset from a large production enterprise network with labeled red team data. It presents 1,648,275,307 events in total for 12,425 users, 17,684 computers, and 62,974 processes. Due to the production nature of the network, this data is heavily anonymized which leads to the loss of some fidelity. For example, authentication is distilled to the following record:

*1,C625$@DOM1,U147@DOM1,C625,C625,Negotiate,Batch,LogOn,Success.*

This is likely a sanitized Kerberos authentication log entry. If this was collected from the unencrypted Kerberos network traffic we are missing the renewability and forwardability of the Kerberos ticket that was issued; however, if this was collected from the domain controller we are missing the event ID as well as the Logon type, two pieces of information that a routinely used by cyber defenders and forensic analysts to determine the source of a cyber incursion [7]. It is worthy to note that LANL also published a dataset in 2017 called the "Unified Host and Network Dataset" [8] which included a much higher level of detail, but did not include any labeled red team activity.

Mordor [9] is another open source project that provides pre-recorded security events generated by simulated adversarial techniques. This data and project is outstanding. The main constraint is scale. Most of the artifacts are for small host-based actions, the only dataset of large enough scale to test with has only two hosts in the environment. This is adequate for most heuristic or signature based testing, but does not provide enough data for many ML based approaches. In addition, LARIAT [10] is a MIT Lincoln Laboratory tool that was initially designed in 2001 as a traffic generator and is used today in several large-scale cyber ranges. It was used to test detection techniques of that era. It has grown from that initial project to possess a suite of capabilities for user and network traffic generation. However, like previously mentioned platforms, it leaves artifacts associated with the framework in both the network as well as the host logs that convolute its use for ML tasks.

III. MOTIVATION FOR PICODOMAIN DESIGN

In this work, we believe that the environment in which to generate the cybersecurity research dataset needed to be diverse enough to provide both the services an average user would use and be realistic enough to contain the systems and platforms that are leveraged by modern attackers. Researching cybersecurity trend data from the past decade revealed that Microsoft Windows environments are decisively the target of attackers worldwide.

At the time of this writing, Windows binaries are submitted to Virus Total, an Alphabet run cloud malware scanning service, a staggering 49.6 time more often than Linux ELF binaries [11]. This is despite a recent survey of hackers worldwide that concluded traditional antivirus (AV) security is irrelevant or obsolete [12]. When specifically asked why AV was obsolete, the respondents stated, "humans are the most responsible for security breaches." This is not surprising. Users must be provided a level of privilege within their environment to accomplish business tasks, this often includes the use of several network resources (Network Share, SharePoint, Email, Windows Active Directory, etc.) as well as specialized software like the Microsoft Office Suite, Adobe Acrobat, and any industry specific tools. This basic level of privilege is often the only level of privilege an attacker needs to gain a foothold within a network.

At the same time, a client-side attack happens when a user is leveraged to download and execute a piece of malware. Such attacks often involve the trojanization of routine and mundane file types, such as Word documents, PDFs, or archives. They are delivered to the victim with a convincing story or realistic guise that tricks these users into running these files. Since these files are legitimate types containing only the smallest necessary amount of malicious code, they often bypass traditional AV detection and prevention methods. Additionally, they either leverage legitimate features of software to gain code execution or possess an exploit for installed software that achieves the same effect.

In summary, one can see that attackers consider Microsoft Windows as the preferred target and client-side attacks as the primary method. In other words, **the most representative attack would be one against a corporation running a Windows environment and the initial foothold attained via a client-side attack.** Therefore, in this paper, this is the scenario used in order to construct the PicoDomain environment in which that data is collected.

IV. PICODOMAIN SIMULATED ENVIRONMENT

The PicoDomain simulated environment consisted of a small Windows-based office environment with five workstations, a domain controller, and a gateway firewall/router. This enterprise was connected to a small-scale internet that housed several websites as well as the adversary's infrastructure.

A. Enterprise Network

The internal network consisted of a Windows Active Directory (AD) environment for the local domain of G.lab. G.lab consisted of three primary Organizational Units (OUs): HR (human resources department), R&D (research and development department), and then a supersecret OU.

- **OU**s are hierarchical entities that allow the segmentation of policies and privileges within Active Directory environments. These policies and procedures can apply to both users as well as machines. In the .lab domain the OUs had the following functions:

- **HR**: Limit HR users' access to only the HR OU's computers and resources

- **R&D**: Permit R&D users to research and conduct experiments without excessive permissions

- **Supersecret**: Greatly restrict access to a single resource

The G.lab network consisted of a single 10.99.99.0/24 network with the domain controller (10.99.99.5) handling DHCP as well as hostname resolution within the network. Due to the DHCP configuration of the network, machine-to-user mappings are best done with hostnames and not IPs. This is a common challenge in DHCP environments, and this dataset is no exception. At the beginning of the scenario the following mapping existed:

*Table 1: Initial User Mapping*

| Hostname | IP | Primary User |
|---|---|---|
| RND-WIN10-2 | 10.99.99.27 | RMOLE |
| RND-WIN10-1 | 10.99.99.29 | JSNAKE |
| HR-WIN7-2 | 10.99.99.30 | BDUCK |
| HR-WIN7-1 | 10.99.99.152 | JDOE |
| SUPERSECRETXP | 10.99.99.160 | SQUIRREL |
| CORP-DC | 10.99.99.5 | AMINISTRATOR |
| PFSENSE | 10.99.99.100 | ROOT |

Routing was handled by a PFSense firewall with an internal IP of 10.99.99.100. This firewall was not joined to the G.lab domain but did manage DNS for the G.lab domain for inbound requests and was the DNS server for all outbound requests. A diagram of this network can be seen in Figure 1.

B. Internet

The internet consisted of the .inet top level domain (TLD). Within this TLD there are six websites: vacation.inet, books.inet, steak.inet, icecream.inet, falcon.inet, design.inet.

Each hosted a modern webpage designed around a single topic. These webservers were Ubuntu 16.04.2 LTS Linux severs running Apache2 with self-signed certificates. Servers were divided among 3 subnets.

Internet traffic was routed by a single router having an interface in each of the 3 different /24 subnets where the simulated internet existed. The router was a PFSense 2.4.3 Virtual Machine (VM).

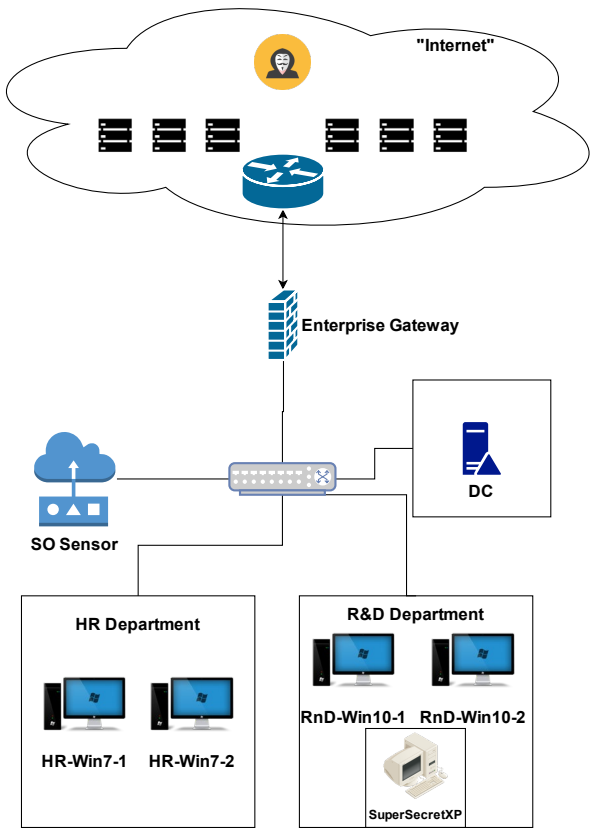

*Figure 1: Network Diagram*

## C. User Emulation

User emulation is often the most challenging portion of any simulation. Any simulated environment should have users that act as real as possible, but this either requires automation or heavy manual interaction. Automation techniques install scripts or agents that leave unwanted traces in collected log data whether from local events that they initiate or from network traffic of the agent interacting with a C2 server. Due to the intended use of the dataset, a hybrid approach was used.

In this work, two scripts were deployed on the endpoints and manual logins via the hypervisor were conducted to generate logon, logoff, and restart actions. Specifically, the scripts that were deployed emulated web browsing and SMB file sharing. These actions were selected as they provide critical services for any small network environment but can also be leveraged by attackers.

## D. Adversary

The adversary consisted of a 2019.1 Kali Linux machine running PowerShell Empire and pivoting traffic via a reverse proxy emulating domain fronting. The domain fronted host was icecream.inet with an IP of 3.3.3.5. The actual Kali Linux machine had an IP of 1.1.1.11.

Domain Fronting is the process of leveraging cloud infrastructure routing mechanisms to obfuscate the actual destination of encrypted traffic. In this scenario, the use of a reverse proxy on a webserver mimics this functionality. This was done via the apache2 mod_rewrite module. If a specific collection of header fields were present in a GET request, the traffic was forwarded to the C2 server; else, the normal website was returned. This process has the benefit of all TLS beginning and ending on the "legitimate" server. The web server decrypts and sends the C2 traffic in plain text to the actual C2 server.

## V. ADVERSARY CAMPAIGN

The adversarial campaign was based heavily off the conclusions reached in Section III above, that an attacker would most likely leverage a client-side attack against an average user to gain the initial foothold. It was crucial to stay within the bounds of this assumption throughout the campaign; however, this assumption did not provide any significant guidance for subsequent phases of the attack lifecycle.

The Mandiant Attack Lifecycle (MAL) [4] was used as the framework upon which to map the adversarial campaign plan. This framework provided the most emphasis on the cyclic nature of key phases of any adversarial campaign. This, in turn, allowed it to be readily matched to forensic artifacts in a chronological manner. The initial phases of the MAL are: Initial Reconnaissance, Initial Compromise, and Establish Foothold. Following this is the cyclic phase of the MAL that is executed as many times as needed prior to reaching an attacker's objective. Within the cyclic phase are four sub-phases: Escalate Privileges, Internal Reconnaissance, Move Laterally, and Maintain Presence. Finally, the MAL concludes with the adversary completing the mission. In the following paragraphs we detail how the adversary progressed through these phases within PicoDomain. Figure 2 provides a visual representation of the domain compromise.

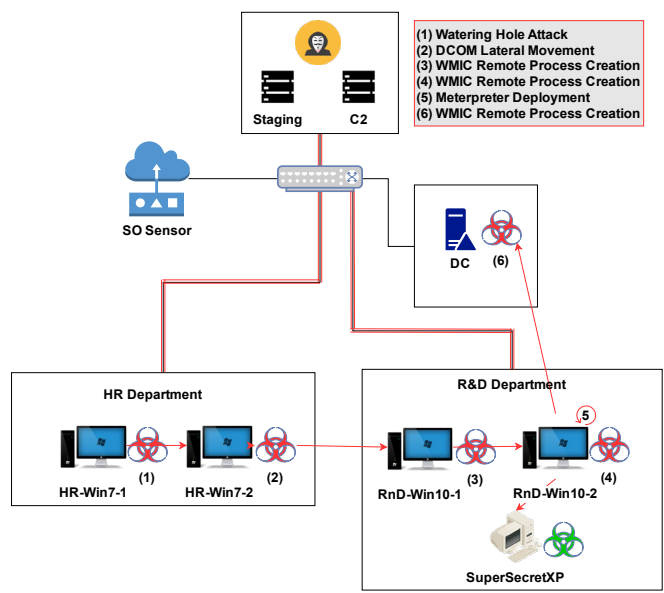

*Figure 2: Attack Diagram*

*A. Initial Compromise*

The initial payload was a dropper via the WinRAR arbitrary file creation (CVE-2018-20250) vulnerability. In this project, the payload was hosted in the guise of an archive for a discount code for a membership service. When extracted, this archive placed a binary in the Startup folder of the current user that executed the next time the user logged in. This binary was a precompiled winexec [13] payload that downloaded and executed the malware to establish the initial foothold in the network.

*B. Establish Foothold*

The malware used to establish the foothold was a semi-custom, three stage design built entirely in PowerShell. The first stage of this malware disabled certificate checking for the downloading of scripts and then downloaded and executed the 2$^{nd}$ stage. By default, PowerShell will not download scripts over HTTPS from sources with self-signed certificates [14]. This script looks innocuous and will not flag on any known signatures. The second stage is an Anti-Malware Scanning Interface (AMSI) bypass based on the rasta-mouse AMSI buffer bypass [15]. This script patches the AMSI process in memory resulting in the AMSI process no longer scanning running scripts. This stage then calls the final stage which is an obfuscated PowerShell Empire launcher. The final stage is obfuscated with Daniel Bohannon's Invoke-Obfuscation script [16].

*C. Escalate Privileges*

Once the foothold on the initial target machine (patient 0) was established, exploitation turned to privilege escalation. Research by Google's Project Zero in 2014 [17] and Foxglove Security in 2016 [18] detailed a new kind of attack that is not easily mitigated due to backwards compatibility. The attack, dubbed "Hot Potato", combines a New Technology LAN Manager (NTLM) relay with NetBIOS Name Service (NBNS) spoofing to execute a command as NT AUTHORITY\SYSTEM, the Windows' equivalent of root. It does this by spoofing the Web Proxy Auto-Discovery (WPAD) IP to 127.0.0.1 and then requesting NTLM authentication to this local server that is adversary controlled.

With command execution as SYSTEM, the compromised user account was made a local administrator. During the next reboot of the system the WinRAR launcher created a new agent with system privileges. Next, credentials were harvested with a PowerShell implementation of Mimikatz.

*D. Maintain Presence (Persistence)*

Persistence was attained through three methods. The first was a second-order effect of the WinRAR exploit. This already deposited an executable in the startup folder of the victim user thus granting persistence that was triggered every time the user logged in. This was only used on the patient zero.

The second form of persistence was via registry keys. Windows relies on registry keys to hold critical pieces of information about the operating system, hardware, and software. One part of the registry controls the applications that are run at startup. In this scenario, a PowerShell Launcher batch script was hidden in the HKCU:SOFTWARE\Microsoft\Windows\CurrentVersion\Run registry key which spawned an agent upon the login of the current user, similar to how the WinRAR based launcher functioned. This technique is typically easy to detect and remediate.

The primary method for persistence was done via WMI event subscriptions. WMI event subscriptions trigger when a series of statements evaluate to true. They are more difficult to find than traditional scheduled task or registry persistence, but increased awareness in the past 4 years has decreased their viability against hardened networks. In this scenario the PowerShell Empire launcher was hidden as a base64 encoded string in the filter "Destination" field. This form of persistence can only be executed with administrative rights.

*E. Lateral movement*

Lateral movement was conducted via WMI and DCOM remote process creation. The DCOM was a PowerShell implementation by Steve Borosh of Matt Nelson's initial technique [19]. The WMI technique was either done via the PowerShell Empire module or via the windows wmic.exe [20].

*F. Complete Mission*

Throughout the campaign, credentials were periodically harvested from each machine, this permitted an administrator account to be compromised when they logged onto a compromise machine. With these credentials an agent was deployed on the domain controller and RDP was enabled. At that time, a meterpreter [21] agent was deployed on a target that was softened by a PowerShell empire agent running as the local administrator. This meterpreter session was then used as a socks proxy to establish an RDP connection to the domain controller in the environment, login with harvested credentials, and then access sensitive documents. This bypassed the need to interact with the cloistered sensitive machine as the sensitive documents were stored in a heavily restricted shared folder within the domain controller itself. Even though the domain admin's direct access to these files was not possible, after modifying the owner of the files and reconfiguring the access permissions, the sensitive documents were able to be exfiltrated.

VI. DATA COLLECTION

Data were collected on a Security Onion 16.04 [22] sensor that received traffic from a SPAN port on the virtual switch to which the entire enterprise was connected.

*A. Red Log*

This file contains a record of the adversarial actions that took place within the environment. It includes the machines involved, targeted user (either the user account used to conduct an action or the user compromised from an action), and the action that occurred. The timestamps were recorded manually and may have variations between them but are accurate within 1 minute of the actual event.

*B. Zeek Logs*

The Zeek logs [3] (formerly known as Bro logs) are from the July 19th 2019 to July 21st 2019. They contain a robust selection of log types used by the Zeek installation on Security Onion 16.04 and would be representative of what would be collected by Zeek sensors in most modern sensing environments. Zeek was configured to log in UTC and save logs as JSON as is the current industry standard. The logs available are the following types: conn, dce_rpc, dhcp, dns, files, http, kerberos, known_hosts, known_serviees, ntlm, pe, smb_files, smb_mapping, software, ssl, weird, and x509.

Below is an example of a Zeek Kerberos log JSON object.

```
{
    "ts": "2019-07-20T12:22:07.237641Z",
    "uid": "CIKD7Hvs7PfUsMBHh",
    "id.orig_h": "10.99.99.152",
    "id.orig_p": 52081,
    "id.resp_h": "10.99.99.5",
    "id.resp_p": 88,
    "request_type": "AS",
    "client": "jdoe/G",
    "service": "krbtgt/G.LAB",
    "success": true,
    "till": "2037-09-13T02:48:05.000000Z",
    "cipher": "aes256-cts-hmac-sha1-96",
    "forwardable": true,
    "renewable": true
}
```

Here we can see the user jdoe sending a request to the Authentication Service (AS) for access to the Kerberos Ticket Granting Ticket (krbtgt). This is indicative of a user's logon to a system. In this log we can see both the source and destination of the communication as well as the unencrypted details of the request. This logon is likely benign as we can see that this authentication event is originating as jdoe from jdoe's main workstation.

The following is an example of a Zeek SSL JSON object that was generated via malicious C2 traffic.

```
{
    "ts": "2019-07-20T18:05:52.328220Z",
    "uid": "Csj4tz91kAqqEUNN8",
    "id.orig_h": "10.99.99.152",
    "id.orig_p": 53825,
    "id.resp_h": "3.3.3.5",
    "id.resp_p": 443,
    "version": "TLSv10",
    "cipher": "TLS_RSA_WITH_AES_128_CBC_SHA",
    "server_name": "icecream.inet",
    "resumed": true,
    "established": true,
    "ja3": "6312930a139fa3ed22b87abb75c16afa",
    "ja3s": "4192c0a946c5bd9b544b4656d9f624a4"
}
```

This entry by itself cannot be demonstrably identified as malicious; however, through its observation in contrast to other Zeek log entries it can be associated with anomalous behavior that can ultimately be linked to malicious events. This is examined in detail in section VII.

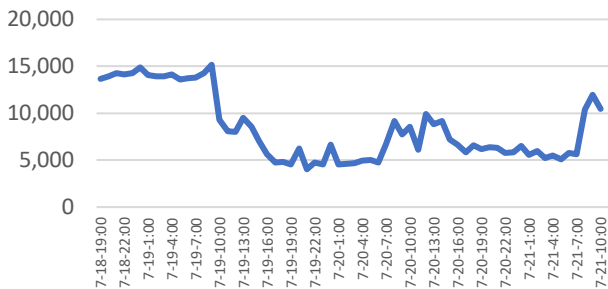

*Figure 3: Log Count Over Time*

Looking at the total amount of logs over time as shown in Figure 3, we can see that an initially high level of activity was followed by two days of normal circadian spikes. This is indicative of the scenario described above. Adversaries have historically identified the human lifecycles of the cyber defenders of a targeted network. They would then use this information to conduct decisive actions when staffing was at its lowest in order to increase their odds of success. This attack starting on a Friday afternoon (July 19[th], 2019) and being executed throughout the weekend (July 20[th]-21[st], 2019) fits this trend.

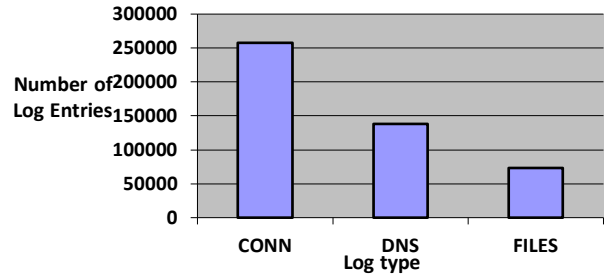

*Figure 4: Log Types*

These logs have undergone sanitization to remove personally identifying information as well as remove references to unpublished works. This process affected approximately 0.2% of log entries, none of which were linked to adversary activity. Figure 4 presents the distribution of three log types.

VII. TESTING

The resulting dataset has been used to refine our ongoing research into the use of machine learning on graph data structures for cybersecurity applications. As mentioned in the introduction, a dataset was needed that was representative of the modern enterprise network, had labeled red team activity, but was also small enough to rapidly iterate prototype algorithms and applications. To that end, the PicoDomain dataset proved to be highly valuable.

A. Event Verification

This section will highlight the presence of the adversarial activity in the logs and demonstrate some potential analysis techniques. This analysis was done through a Jupyter notebook using the red team log and statistical analysis.

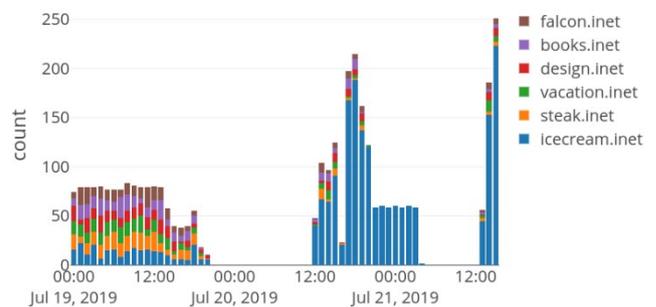

*Figure 5: SSL Activity by Domain*

When examining the SSL log activity in Figure 5 we can see a clear distinction between behavior before and during the adversary activity. On Friday July 19[th] we see a near even distribution of traffic between the sites on the simulated internet. After normal activity concluded on the 19[th] we see the SSL activity subside, and then when activities resume on Saturday July 20th, there is a clear increase in the quantity of connections to the domain being used for C2. This type of behavior deviation could be used by analytics to detect compromised systems and/or malicious domains.

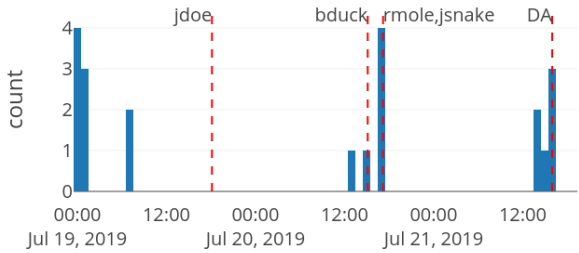

*Figure 6: New Kerberos Activity*

In Figure 6 we see the frequency of new Kerberose authentication events as extracted from the Zeek Kerberos logs. A new Kerberos authentication event is defined as the occurance of a unique bigram extracted from the Kerberos logs between either a source and destination IP or a Kerberos client and service. An established network should produce unique bigrams in relatively few situations, such as when a new user is made or when a user changes roles. In Figure 6 we have overlayed the points in time where each account was compromised. With the exception of the initial compromise, there is tight correlation between compromised accounts and spikes of new Kerberos authentication activity. These Kerberos bigram events could be utilized to detect compromised accounts.

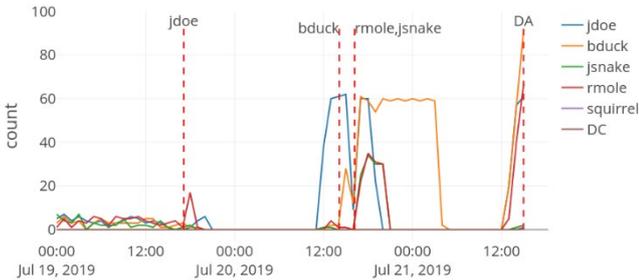

*Figure 7: SSL Activity to C2 Domain*

Another method for detection of the adversarial C2 traffic within the datset was to examine when each user's primary work computer began to increase its communication to the C2 domain. In Figure 7 we again see a strong correlation between the time when each account was compromised and that user's primary machine increasing the frequency with which it communicated to the C2 domain. This could be used in conjunction with the previous analytics to identifiy compromised users.

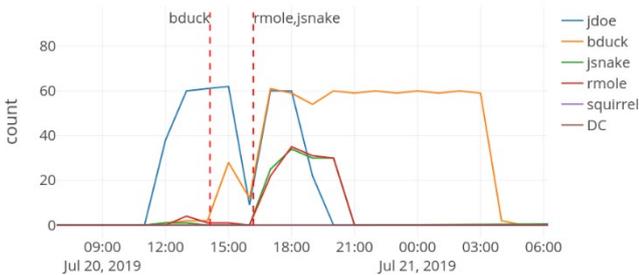

*Figure 8: SSL Activity to C2 Domain (Zoom-In)*

Focusing on the 2nd day of data, the most active day within the data as shown in Figure 8, SSL activity by user to the C2 domain further demonstrates correlation between the compromise of user bduck and then shortly thereafter rmole and jsnake. It is also worth mentioning that this figure shows the adversarial communication that preceeded the compromise of new users in the blue spike of jdoe.

### B. Unsupervised Machine Learning

To demonstrate that this dataset contains sufficient information for machine learning analysis, we utilized an off-the-shelf anomaly detection algorithm from the sklearn Python library. Specifically, we implemented the Local Outlier Factor (LOF) originally proposed in the paper LOF: identifying density-based local outliers [23]. This is a relatively straight forward unsupervised anomaly detection algorithm based on identifying points with significantly different densities than their k-nearest neighbors. We extracted features from the Kerberos authentication logs via a 1-hot encoding of the client principal and service principal. Table 2 shows the anomalous authentication events as reported by the LOF technique. We can see that 10 out of the 14 reported events are true positives while the remainder are false positives. Most of these events would not be detected by traditional signatures. Additionally, heuristic detection mechanisms that would detect them would need to be modified, likely manually, for each network they are applied to. For example, an office without specified work spaces would have a very different physical machine to user account mapping then a traditional office with assigned work spaces. A heuristic designed for the latter would produce voluminous false positives in the former. This illustrates the strength of an unsupervised learning approach, and demonstrates the utility of the PicoDomain dataset for developing and prototyping such techniques.

*Table 2: Pico Domain Unsupervised ML Anomalies*

| Auth from | Auth to | Notes |
|---|---|---|
| jdoe | rpcss/hr-win7-2 | True Positive |
| jdoe | bduck | True Positive |
| rmole | host/hr-win7-2 | False Positive |
| rmole | protectedstorage/corp-dc.g.lab | False Positive |
| bduck | hr-win7-2 | False Positive |
| bduck | host/rnd-win10-2 | True Positive |
| bduck | hr-win7-2 | False Positive |
| rnd-win10-2 | gc/corp-dc | True Positive |
| rnd-win10-1 | gc/corp-dc | True Positive |
| jdoe | host/rnd-win10-2 | True Positive |
| jdoe | host/rnd-win10-1 | True Positive |
| bduck | host/rnd-win10-2 | True Positive |
| local.admin | krbtgt | True Positive |
| local.admin | krbtgt | True Positive |

### C. Runtime Comparison

To show how valuable the PicoDomain is in terms of its size and ease of use, we took the same off-the-shelf machine learning technique discussed previously, and this time applied it to an existing labeled cybesecurity dataset. We utilized the Los Alamos National Labs (LANL) Comprehensive Multi-source Cyber Security dataset [2]. This is one of the few datasets which contains network level events, and labeled red team activity. However, as mentioned previously, this dataset poses difficulties as it is highly anonymized, and highly verbose, making it a challenging dataset to develop algorithms on. Table 3 shows some metrics for various stages of the algorithm pipeline. We can see that in all cases, the PicoDomain (half a million logs) is significantly less burdensome to utilize, with the entire algorithm pipeline completing in only a few seconds, and utilizing several kilobytes of memory, vs the LANL pipeline which takes over 2.5 hours, and 70 GB of memory. Although the LANL dataset is very useful for testing algorithms, its size (1.6 billion logs), especially sparsity of labels and anonymization, make it difficult to use for algorithm development and prototyping. In our research, we chose to first quickly iterate our algorithms in the PicoDomain dataset, followed by the evaluation on the LANL dataset. We believe such an approach can also be very beneficial to other researchers in this area.

*Table 3: Time and memory comparison PicoDomain vs. LANL 2015*

| Metric | LANL | Pico |
|---|---|---|
| Parsing raw logs into Pandas Dataframe time | 1699 s | 0.2 s |
| Dataframe memory usage | 70 GB | 59 kB |
| Data manipulation & feature extraction time | 1834 s | 1.3 s |
| Local Outlier Factor anomaly detection time | 6835 | 0.09 |

## VIII. FUTURE WORK

This dataset met an immediate need for documented and representative data for experimentation but has several areas in which future work can enhance both its fidelity and its robustness.

### A. New C2 Platforms

PowerShell Empire is, unfortunately, no longer being actively developed. It was an exceptional project, bringing together a comprehensive toolkit for almost every phase of an attack; however, the rise of awareness around PowerShell based attacks both in the IT sector as well as within Microsoft has diminished its usefulness since its initial release. At the same time, as PowerShell was losing its dominance as an adversary platform, researchers started experimenting with .NET based malware. Recently Ryan Cobb of Specter Ops release Covenant [24], a collaborative and extensible C2 framework and platform based exclusively on .NET. As detection methodology for .NET malware is in its infancy, its use will likely follow the arc of PowerShell as an adversary tactic and remain relevant for the next several years.

Several other C2 platforms, e.g., PoshC2 [25], FactionC2 [26] and MerlinC2 [27], exist that decouple the management of communication and the employment of code on the target endpoints, focusing more closely on being a framework than an all-in-one solution. Utilizing one of these platforms with custom agent code would present a greater challenge for both traditional and Machine Learning based detection methods.

### B. Scope & Scale

The enterprise presented in this scenario is very similar to something that would be found in a small business. Although the traffic captured qualitatively looks like something you would find in large networks the scale is not there. The limitation in scope for this endeavor was based on the need to prevent simulation tool artifacts from tainting the logs.

Future work will consist of investigating the possibility of effective log cleansing or the use of hypervisors to automate user interaction. The former being smaller in scope but the latter providing results that are forensically indistinguishable from an interactive user. Leveraging either technique would allow the simulated environment to be scaled up dramatically, likely only being limited by hardware availably.

### C. Host Logging

This dataset was focused exclusively on network traffic and the industry standard logs that are generated from it. This was done due to the research it was supporting, but also from the likelihood of network data being available to security professionals in the event of an incident. Centralized host log collection of any efficacy is unlikely to be configured within a network not running a professionally installed Security Information Event Management (SIEM) system. If not configured prior to an incident, it is unlikely to be configured in the aftermath of an attack. The configuration of forwarding of logs puts a large burden on network operators and the volume of logs being forwarded may saturate network links. The network data, specifically Zeek, can be gathered by configuring SPAN ports or deploying TAPs. Both of these processes occur passively and do not affect network operations but for the moment a TAP is turned on. This makes the barrier to entry relatively low even for an understaffed network operations department. Additionally, the nature of collection allows incident responders to bring in their own equipment after an event and rapidly generate data.

Despite the availability of network data, it misses large amounts of details in the event of a cyber-attack. For example, when deploying WMI event subscription persistence the network only saw the HTTPS connection. At this same time, if event logging was properly configured, one would see the minutia of the WMI event and even the encoded launcher. This would allow the user to rapidly decode the address of the C2 server and respond accordingly. This level of detail should be captured in future work.

### D. Improved Cycle Representation

Certain log types exhibit no activity at night. This is an unfortunate area where the dataset does diverge from real-world data. In a traditional office network, the amount of HTTP traffic at night is expected to be several times lower than during the day, perhaps as low as 1% of normal daily traffic, but not nonexistent. This nighttime traffic typically consists of updates, backups, and 24/7 system services. These are all things that do not rely on user interaction. After analyzing the logs it becomes apparent that these services are still active by the number of DNS queries during the nighttime hours, but these queries are for internet based services such as Windows Update. As these services were not deployed in the simulated environment the DNS queries do not resolve to reachable IP addresses. Future work would strive to correct this imbalance.

## IX. CONCLUSION

Existing network-level cyber security datasets are either lacking in ground truth, lacking in completeness, or not representative of real-world scenarios, rendering them insufficient for algorithm development and prototyping. In this work we generated the PicoDomain dataset to address these shortcomings. This dataset provides researchers with a compact, yet complete representation of an enterprise computer network, with an included and labeled complete attack campaign. This dataset is useful, as well as easy-to-use, for not only simple statistical analysis, but also more complex machine learning tasks.